\documentclass[]{spie}  

\usepackage{amsmath,amsfonts,amssymb}
\usepackage{graphicx}
\usepackage[colorlinks=true, allcolors=blue]{hyperref}
\usepackage{afterpage,placeins}

\newcommand\arcsec{$^{\prime\prime}$}

\title{Turbulence nowcast for the Cerro Paranal and Cerro Armazones observatory sites}

\author[a]{J. Milli}
\author[b,c]{T. Rojas}
\author[b]{B. Courtney-Barrer}
\author[b]{F. Bian}
\author[b]{J. Navarrete}
\author[d]{F. Kerber}
\author[b]{A. Otarola}

\affil[a]{Universit\'e Grenoble Alpes, IPAG, F-38000 Grenoble, France }
\affil[b]{European Southern Observatory, Alonso de C\'ordova 3107, Casilla 19001, Santiago, Chile}
\affil[c]{Dept. of Electrical Engineering, Universidad de Chile, Santiago, Chile}
\affil[d]{European Southern Observatory, Karl-Schwarzschild-Stra{\ss}e 2, 85748 Garching, Germany}

\authorinfo{Further author information: Julien Milli: E-mail: julien.milli@univ-grenoble-alpes.fr}

\begin{document} 
\maketitle

\begin{abstract}

Optical turbulence affects significantly the quality of ground-based astronomical observations. An accurate and reliable forecast of optical turbulence can help to optimize the scheduling of the science observations and to improve both the quality of the data and the scientific productivity of the observatory. However, forecasts of the turbulence to a level of accuracy that is useful in the operations of large observatories are notoriously difficult to obtain. Several routes have been investigated, from detailed physical modelling of the atmosphere to empirical data-driven approaches. Here, we present an empirical approach exploiting spatial diversity and based on simultaneous measurements between two nearby sites, Cerro Paranal, host of the Very Large Telescope (VLT), and Cerro Armazones, future host of the Extremely Large Telescope (ELT) in Chile. We study the correlation between the high-altitude turbulence as measured between those two sites. This is part of the on-going efforts initiated by the European Southern Observatory (ESO), to obtain short-term forecasts of the turbulence to facilitate the operations of the VLT and prepare the ELT mode of operations.

\end{abstract}


\section{Introduction}
\label{sec_intro}  

\subsection{The value of turbulence nowcast}

A nowcast represents an extrapolation of the current conditions to the very near future, typically a few hours. This term comes from the contraction of ``now'' and ``forecasting''. It is widely used in meteorology to include details that cannot be solved by numerical weather predictions (NWP) that capture the physics of the phenomenon and can forecast longer time scales but often have a limited spatial and temporal resolution. It is therefore highly complementary to standard NWP. In addition, nowcasts are typically fast to obtain and straightforward to deploy, unlike NWP which require significant computing resources to run hour-long simulations.

Both approaches have been combined in the context of mesoscale numerical models, where the model output is filtered using, for instance, autoregressive techniques or Kalman filters  (Masciadri et al. 2020\cite{Masciadri2020} , Turchi et al. 2019 \cite{Turchi2020}). This showed an improvement in the accuracy of the seeing predictions at 1h for the Large Binocular Telescope (LBT): while the accuracy as measured by the seeing root mean square error (RMSE) is 0.30\arcsec{} with standard NWP, it reaches 0.11\arcsec{} when combined with an autoregressive filter. 
Standard NWP techniques achieve a similar level of accuracy at Mauna Kea, with an 
RMSE below 0.25\arcsec{} when the seeing is below 1\arcsec{} (Lyman et al. 2020\cite{Lyman2020}).

The motivations to obtain such short-term predictions were reviewed in detail in Milli et al. (2019\cite{Milli2019_AO4ELT} , hereafter M19) and we only summarise here their main conclusions. 
Service mode, also referred to as queue observing, is the most common operational model of large observatories and rely on the ability of the support astronomer to select and execute the highest-ranked observation compatible with the current atmospheric and turbulence conditions. With the generalisation of adaptive optics (hereafter AO) systems, the instrument performance is now heavily dependent on the turbulence conditions. Therefore, the ability to predict whether the data quality will be sufficient to meet the science goal, also depends on the accuracy of the turbulence prediction over the length of the observation, typically $\sim1$h at the VLT. The performance of current and future AO systems depend upon many  parameters such as the integrated seeing over the line of sight, but also the coherence time of the atmosphere (e.g. Cantalloube et al. 2020\cite{Cantalloube2020}), the fraction of turbulence in the ground layer (e.g. Fusco et al. 2020\cite{Fusco2020}), or for more complex tomographic AO systems the full vertical distribution of the refractive index structure constant $C_n^2$ (e.g. Farley et al. 2020\cite{Farley2020}). Predictions are therefore highly desirable to guide the support astronomer in his/her choice, as well as to optimise the science return of the observatory at large. 
Currently, at the VLT, a study done over one semester for one telescope reports that 4\% of the time spent in service mode is lost because of unsatisfactory seeing conditions\cite{Milli2019_AO4ELT}. More importantly, the programs suffering most are those requesting the best conditions (typically a seeing better than 0.6\arcsec), with a non-compliance rate higher than 10\% in those cases\cite{Milli2019_AO4ELT} . These statistics can generate and feed conservatism, both from the support astronomer who can be tempted to execute low-risk, less-demanding observations to avoid drifting out of the constraints, but also from the community of astronomers who might be reluctant to ask for the best conditions available at the astronomical sites (see Fig. 1 of M19 for instance). 

The future instrumentation of giant segmented telescopes will include all kind of AO systems:  ground-layer, tomographic, multi-conjugated, multi-object and extreme AO systems, each having different behaviours to the turbulence. On the ELT, observations will likely be decomposed into the smallest possible blocks that can be calibrated, so that observations can be interrupted between those blocks in case of adverse conditions. The need for short-term turbulence predictions is therefore very desirable. The VLT is a good environment to implement such a system, adapt the operational tools to this new information and get ready for the ELT. In addition, the pressure for telescope time will be higher on giant telescopes than on 10m-class telescopes, encouraging to optimise science return and reduce time losses even further.

\subsection{Limitations of current nowcast implementations}
\label{sec_limitations}

\begin{figure} [h]
   \begin{center}
   \includegraphics[width=1\hsize]{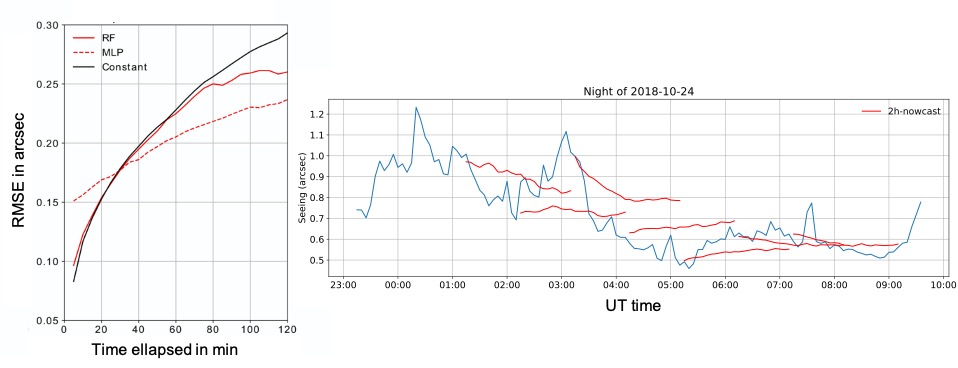}
	\end{center}
   \caption[ex] 
   {\label{fig_ML_nowcast}
   Left: error in the prediction of the seeing as a function of the elapsed time since prediction using two different algorithms: random forests (RF) and a multilayer perceptron (MLP). We overplotted the error obtained in a scenario called ”Constant” where we assumed the seeing is constant over 2h, and equal to the average of the previous 15min. Right: example of 2h-nowcasts for one night, done with the RF algorithm. The blue line shows the seeing as measured by the DIMM seeing monitor. For clarity only a few nowcasts spaced by 1h have been plotted. Both graphs have been reproduced from M19.
  }
   \end{figure}

In M19, we presented a simple empirical approach of turbulence nowcast, based on measurements delivered from the Paranal Astronomical Site Monitor (ASM), the suite of atmospheric and turbulence sensors available at the Paranal Observatory. This method used measurements from 2h in the past, in a machine learning framework, to predict the future evolution of the seeing up to 2h. The accuracy obtained is summarised in Fig. \ref{fig_ML_nowcast} (left). The RMSE is 0.22\arcsec{} for the random forests algorithm and 0.20\arcsec{} for the multi-layer perception algorithm, two off-the-shelf machine learning algorithms we applied in M19. Although this may seem promising, this is only slightly better than the baseline scenario where the seeing is assumed to remain at a constant value, equal to the average seeing over the past 15min (black line). By inspecting individual nights (an example is provided in Fig. \ref{fig_ML_nowcast} right and several more can be found in M19), we see that the general trend is well captured by the algorithm, hence the improvement with respect to the constant scenario for timescales between 1h and 2h. However, sharp fluctuations of the seeing are not captured. For instance, any sudden increase in seeing is systematically missed by the nowcast, which explains why nowcasts provide only a marginal improvement with respect to the constant scenario.  

To overcome this limitation, we proposed in M19 to explore several new paths. First we proposed to employ algorithms better tuned to predict stochastic time series. This path is currently being investigated internally at ESO. However, if the information on the sudden changes in the seeing is not contained in the input data used to train the algorithm, this approach cannot succeed. In M19, we mostly used local data collected from the Paranal platform, such as temperature, relative humidity, pressure, complemented by two integrated quantities over the whole air column, the seeing and the precipitable water vapour.  More measurements from high altitude layers can be useful to better capture the evolution of the turbulence in altitude, such as temperature and humidity vertical profile provided by the Low Humidity And Temperature Profiling Radiometer (LHATPRO\cite{Kerber2012}), or turbulence $C_n^2$ profiles as provided by the Multi-Aperture Scintillation Sensor (MASS\cite{Kornilov2003}) or the SLOpe Detection And Ranging (SLODAR\cite{Butterley2020}) instrument. 

To better deal with sudden changes in the turbulence, we also proposed to include spatial diversity, in order to rely on both temporal and spatial measurements for the prediction, as already proposed in Navarrete et al. 2011\cite{Navarrete2011}. Ideally, a grid of weather and turbulence monitors distributed around the observatory could detect rapid changes in the turbulence before it affects the Paranal Observatory. This grid does not exist yet, but an array of atmospheric monitors is foreseen in the context of the Cherenkov Telescope Array (CTA\cite{Hofmann2017}) that will be located in the valley between Cerro Paranal and Cerro Armazones (see Fig. \ref{fig_config}), and three peaks surrounding Cerro Paranal have been equipped in the past with atmospheric and turbulence sensors in the context of site monitoring campaigns (Cerro Armazones, Cerro Ventarrones and Cerro Vicu\~na MacKenna located East and North-East of Cerro Paranal). In M19, we illustrated the potential of this technique by comparing the seeing measurements on the same night at Cerro Paranal and Cerro Armazones. A burst in seeing is detected, first in Paranal, and 70min later in Armazones. Considering the profile of the wind speed and direction for that night, this burst was compatible with a turbulent layer located at about 4km in altitude, which would have travelled from Paranal towards Armazones, assuming a frozen-flow hypothesis. This interesting case prompted us to investigate more systematically the correlation between the turbulence as measured between Armazones and Paranal, two sites for which there are simultaneous measurements over several years. 

\section{Method and results}

\subsection{Configuration of the sites  and scope of the study}

\begin{figure} [h]
   \begin{center}
   \includegraphics[width=1\hsize]{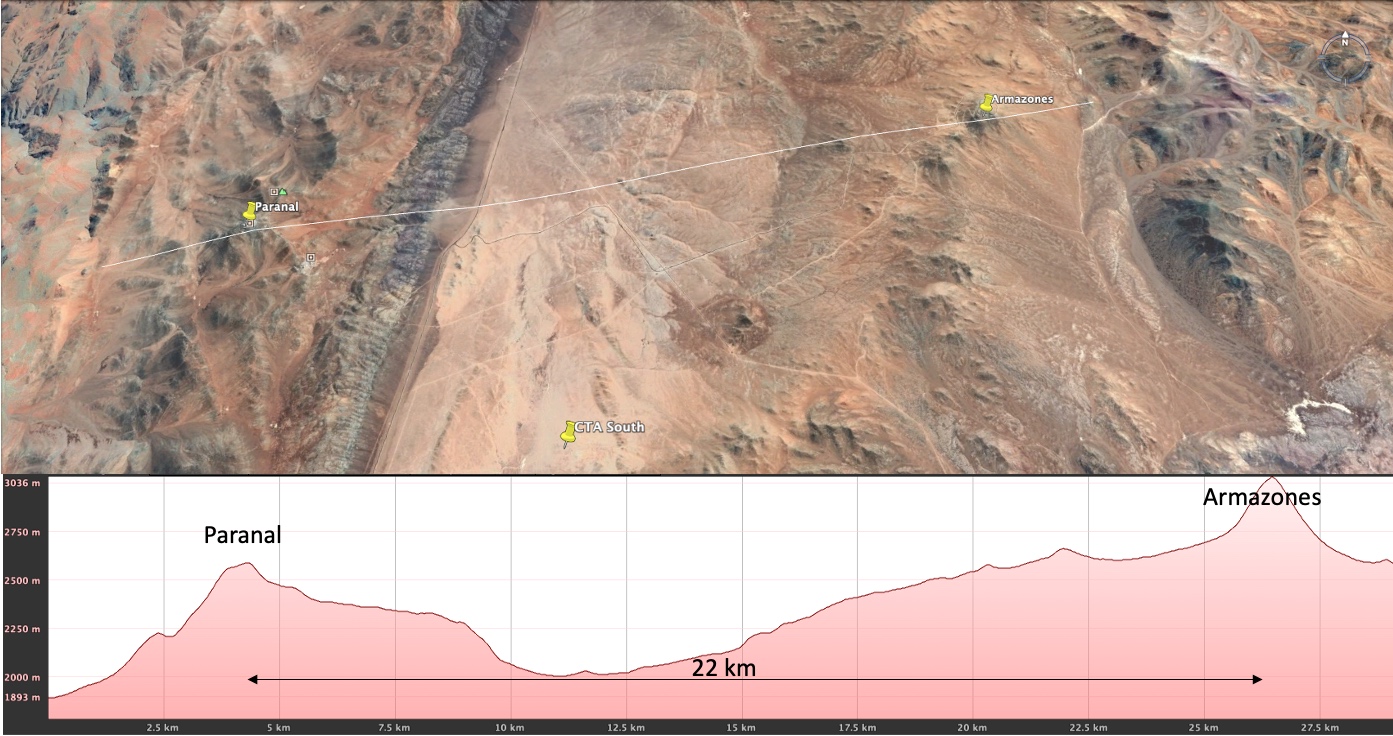}
	\end{center}
   \caption[ex] 
   {\label{fig_config}
Satellite view of the Paranal Armazones area, also indicating the location of CTA South. The North is up. The white line is joining Paranal and Armazones. It has a bearing of $79^\circ$ and its altitude is shown below. Source: Google Earth (credit: 2020 CNES /Airbus)
  }
   \end{figure}

Cerro Paranal and Cerro Armazones are two different peaks 22\,km apart and of altitude 2635m and 3048m respectively, as shown in Fig. \ref{fig_config}. In between those two peaks, there is a valley at an altitude of about 2000m. 

\begin{figure} [h]
   \begin{center}
   \includegraphics[width=1\hsize]{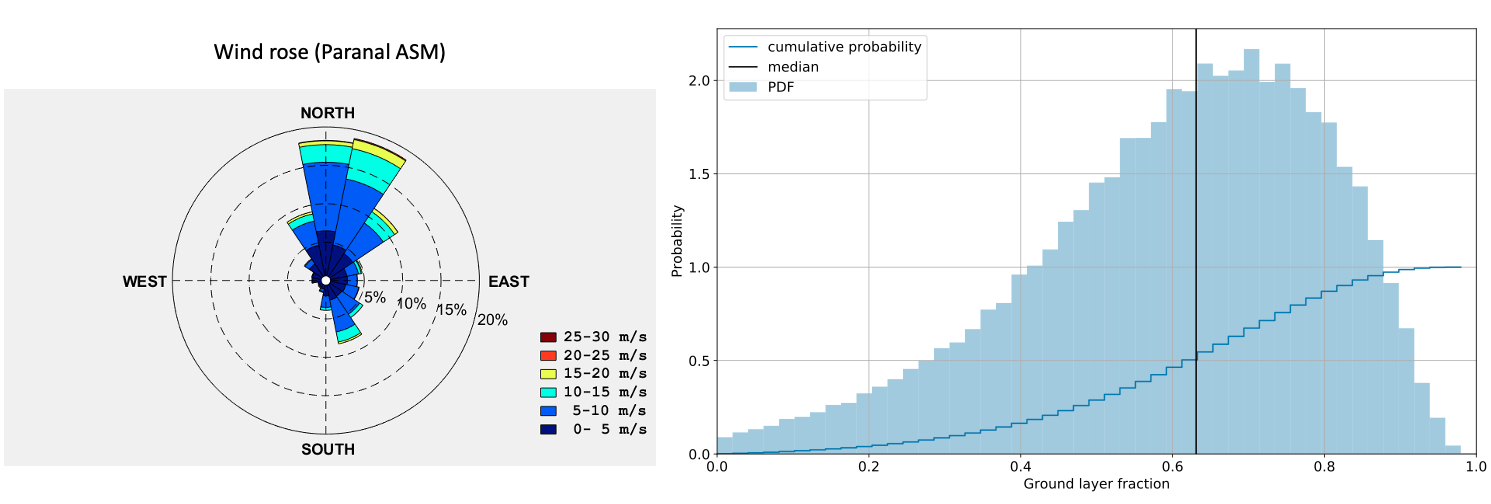}
	\end{center}
   \caption[ex] 
   {\label{fig_wind_rose_GLF}
Left: distribution of the wind as measured from the 30m tower of the ASM at Cerro Paranal at night between 2017 and 2018. Right: probability density function (PDF, filled area) and cumulative probability of the ground-layer fraction, as measured from the MASS-DIMM in Paranal between 2016 and 2019. The median ground-layer fraction of 63\% is indicated as a vertical black line.
 }
 \end{figure}

To validate the approach of temporal nowcast described in the previous section, we studied first the correlation between the integrated seeing, as measured during the same nights between 2004 and 2009 from Paranal and Armazones. We used seeing measurements from the Differential Image Motion Monitor (DIMM\cite{Sarazin1990}). The Paranal DIMM measurements are described and available in the ESO archive of the Paranal ambient conditions\footnote{\href{http://archive.eso.org/wdb/wdb/asm/historical_ambient_paranal/form}{http://archive.eso.org/wdb/wdb/asm/historical\_ambient\_paranal/form}} while the Armazones DIMM data was collected during the Thirty Meter Telescope (TMT) site testing campaign\footnote{The data are presented and available at \href{https://sitedata.tmt.org}{https://sitedata.tmt.org}}.
However, from the thousand nights investigated with simultaneous Paranal and Armazones data available, only a small fraction of them (about 2\%) showed a significant correlation between the integrated seeing measured on both sites. Several reasons can be put forward to explain this lack of significant correlation: 
\begin{itemize}
\item the seeing represents the turbulence strength integrated along the line of sight. It contains therefore the contribution of various atmospheric layers, which travel in different directions and at different speeds with the wind. The correlation that might exist for one given layer is therefore diluted by the contribution of other layers. 
\item a significant fraction of the turbulence comes from the contribution of the ground layer. The histogram and cumulative distribution of the ground layer is shown in Fig. \ref{fig_wind_rose_GLF} (right), as measured from the MASS-DIMM\cite{Kornilov2003} instrument in Paranal. The median value is 63\%, indicating that in more than half the nights, the fraction of the turbulence below $\sim500$m is  greater than 63\%. However, due to the difference in altitude of about 400m between Paranal and Armazones (see Fig. \ref{fig_config} bottom), this ground layer might be significantly different between the two sites, and dominated by local effects due to their corresponding and distinct topography. 
\item the wind in Paranal at night is essentially coming from the North, while the line joining Paranal to Armazones has a bearing of $79^\circ$  ($0^\circ$ means North and $90^\circ$ is East), as shown in the wind rose displayed in Fig. \ref{fig_wind_rose_GLF} (left). Even if the ground layer turbulence is transported without significant evolution by the wind, there are therefore very few cases (a few percents) where it blows in the Paranal-Armazones direction. 

We therefore decided to restrict the scope of the study to study layers where we expect a higher correlation between the two sites. 

\end{itemize}

\subsection{Description of the approach}

Atmospheric and turbulence parameters on Cerro Armazones were monitored from 2004 to 2009 during the TMT site testing campaign (Sch{\"o}ck et al. 2009\cite{Schoeck2009}\footnote{The data are presented and available at \href{https://sitedata.tmt.org}{https://sitedata.tmt.org}}). The data set contains not only the integrated seeing measured by a DIMM, but also low resolution profiles of the turbulence measured by a MASS. Fortunately in September 2004, a light and optimized version of a MASS, called the MASS-LITE instrument,  was installed in Paranal\footnote{The data are presented and available at \href{http://www.eso.org/gen-fac/pubs/astclim/paranal/asm/mass/MASS-LITE/}{http://www.eso.org/gen-fac/pubs/astclim/paranal/asm/mass/MASS-LITE/}}.

   \begin{figure} [h]
   \begin{center}
   \includegraphics[width=0.5\hsize]{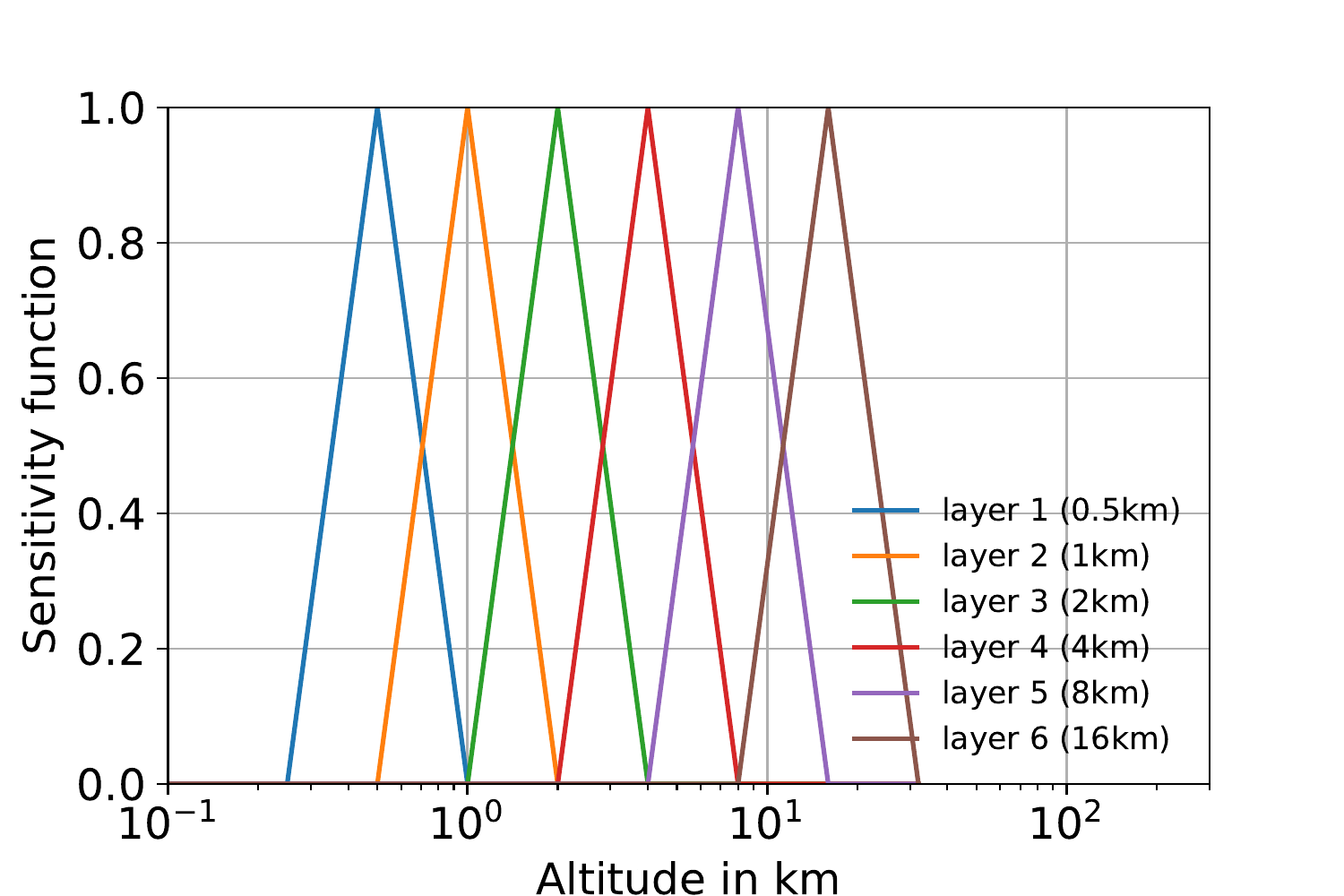}
	\end{center}
   \caption[ex] 
   {\label{fig_MASS_response}
MASS sensitivity functions normalised to 1, reproduced from Lombardi et al. (2016\cite{Lombardi2016}).
 }
 \end{figure} 

\begin{figure} [h]
   \begin{center}
   \includegraphics[width=1\hsize]{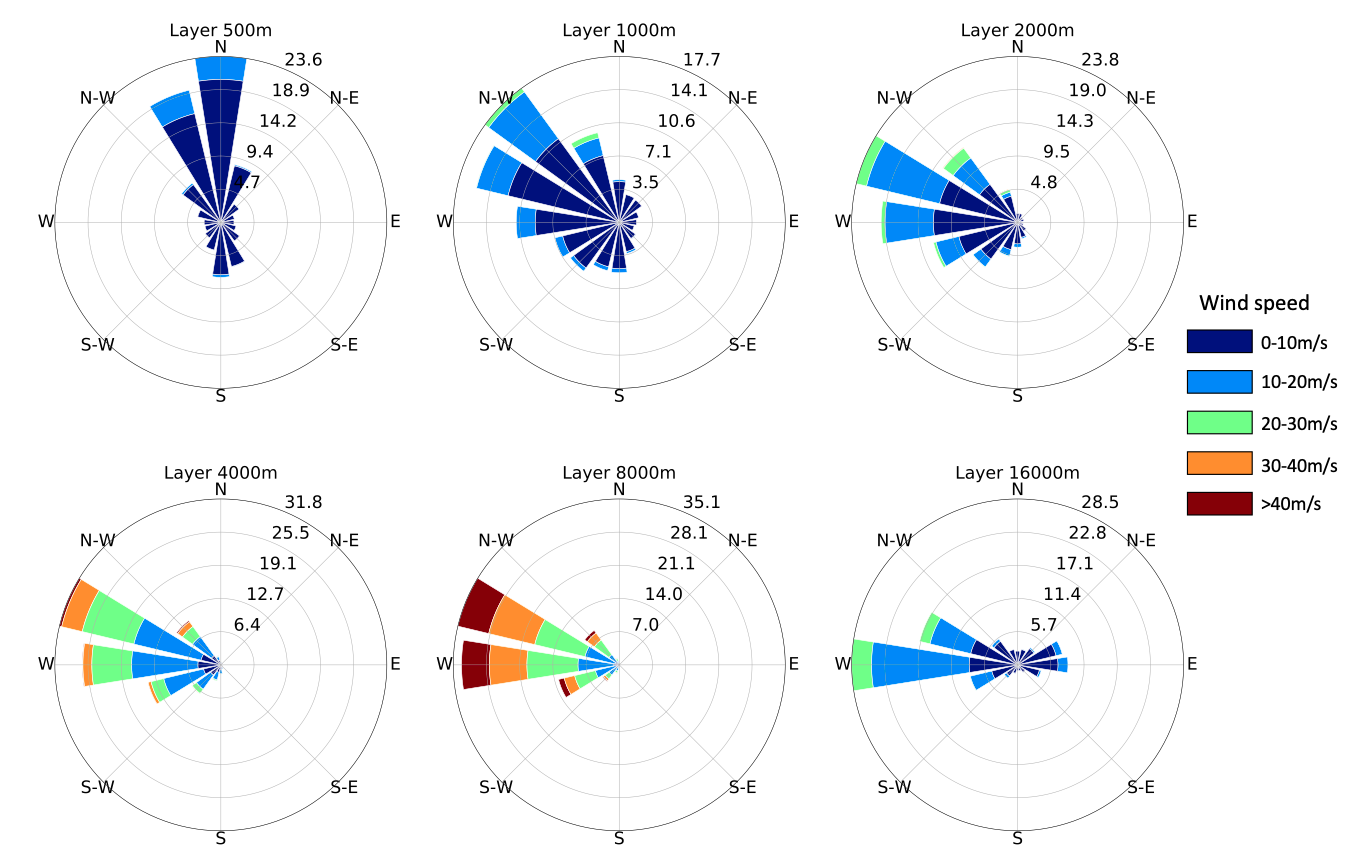}
	\end{center}
   \caption[ex] 
   {\label{fig_wind_rose_MASS_layers}
Distribution of the wind in the six different layers seen by the MASS. The profiles of the wind speed and direction were extracted from global circulation models interpolated at the location of Cerro Paranal provided by the ERA5 reanalysis data\cite{ERA5} from 2004 to 2009, weighted by the MASS response function. The colour represents the wind speed. The wind direction was binned in sectors $22.5^\circ$ wide. The extent of each sector is proportional to the occurence rate in percent for the wind direction to be within this sector. 
 }
 \end{figure}

The MASS instrument provides the turbulence ($C_n^2.dh$ in $m^{1/3}$) in six layers centred at fixed altitudes of 0.5, 1, 2, 4, 8 and 16\,km. The sensitivity of each MASS layer to the turbulence can be simplistically assumed to be triangular, with a peak sensitivity at the layer central altitude (Lombardi et al. 2016\cite{Lombardi2016}). We reproduced the sensitivity functions of each layer in Fig. \ref{fig_MASS_response}. We then applied this sensitivity function to the wind vertical profiles in order to estimate the weighted-average wind speed and direction in the layers probed by the MASS at Paranal and Armazones. We used wind profiles estimated hourly and retrieved from the Copernicus Climate Change Service. It used the fifth generation ECMWF (European Center for Medium Range Weather Forecast) reanalysis for the global climate and weather for the past 4 to 7 decades ((ERA5\cite{ERA5}), interpolated at the Paranal location. We assumed the MASS was located at the height of Paranal, and therefore the central altitudes of the six layers correspond to $\sim3.1$, 3.6, 4.6, 6.6, 10.6 and 18.6\,km above sea level. The resulting distributions are shown in Fig. \ref{fig_wind_rose_MASS_layers}.

The first layer shows a strong contribution of northern winds, similar to the Paranal ground wind distribution shown in Fig. \ref{fig_wind_rose_GLF} (left). However, the wind changes direction in higher layers to come mostly from the West. The jet stream layer responsible for fast westerly wind at a pressure level of about 200\,mbar is clearly detected in the MASS fifth layer (centred at 8\,km above Paranal, or 10.6km above sea level), with winds greater than 30m/s more than 50\% of the time. 

This layer is therefore promising to investigate the correlation between the turbulence at Armazones and Paranal. The high wind speeds would correspond to travel times between 5 and 20\,min (for wind speeds of 70 and 18\,m/s respectively), a time scale short enough for the turbulence not to have significantly evolved between the two sites (frozen-flow hypothesis). Unlike lower layers, the 400m altitude difference between Paranal and Armazones is negligible in the MASS 8\,km sensitivity function.  We therefore restricted the analysis to the MASS 8\,km layer. 

   \begin{figure} [h]
   \begin{center}
   \includegraphics[width=0.9\hsize]{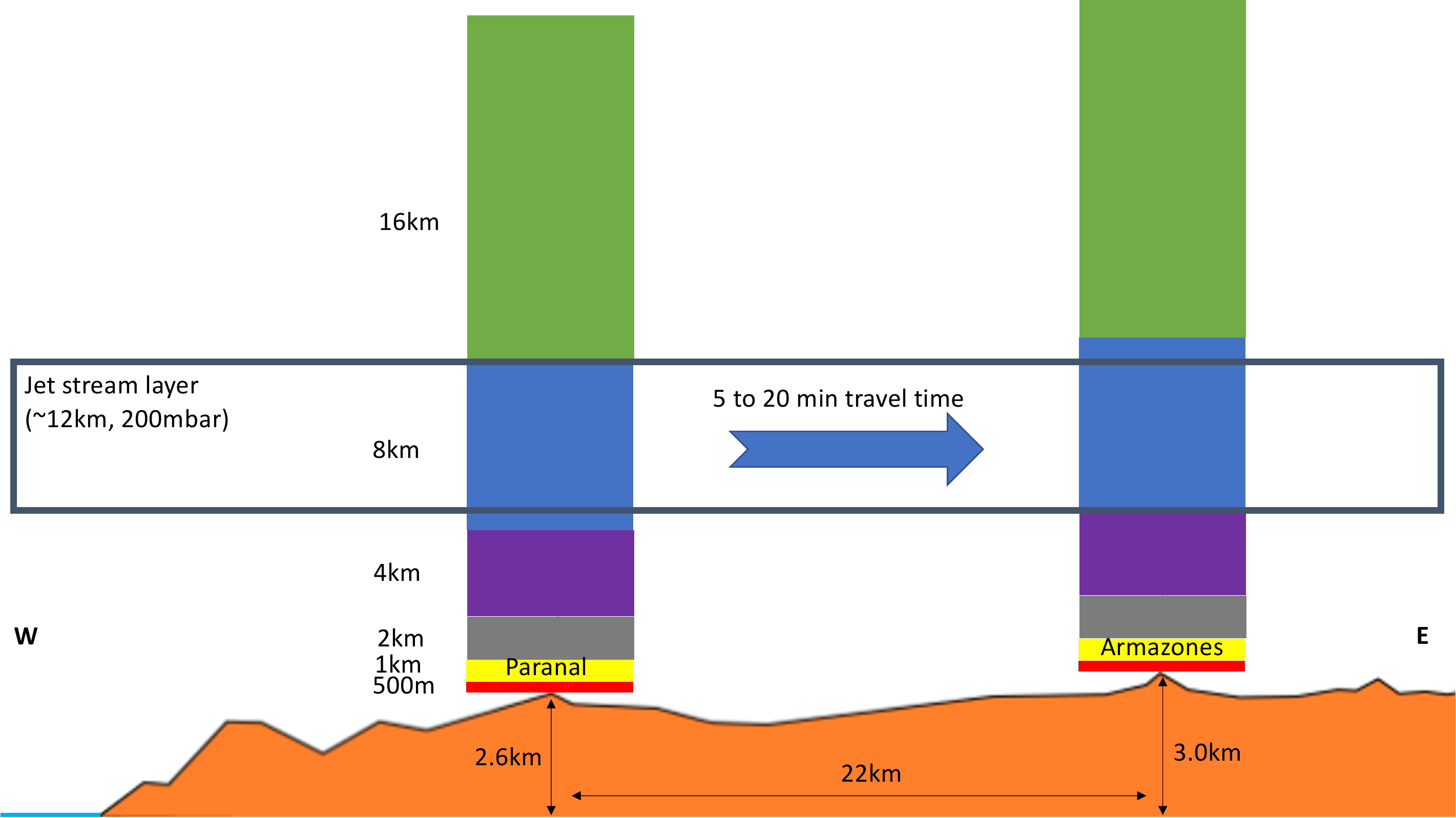}
	\end{center}
   \caption[ex] 
   {\label{fig_cut_MASS_profile}
Sketch of the configuration, highlighting the MASS layer at 8\,km (in blue) which has been selected for this analysis. 
 }
 \end{figure} 

The sampling of the MASS is one measurement every 70 to 80\,s for both instruments. In order to compute the cross-correlation, we resampled both time series regularly every minute. If, for a given time stamp, there was no measurement within one minute, we did not interpolate and left a gap in the time series. 

After selecting the nights with simultaneous measurements from Paranal and Armazones, and with gaps representing less than 10\% of the entire night, we ended up with 571 valid nights, for which we computed the cross-correlation. We used the Spearman cross-correlation because we did not expect necessarily a linear correlation. We however also tried the Pearson correlation and obtained the same conclusions. 

\subsection{Results}

\begin{figure} [h]
   \begin{center}
   \includegraphics[width=0.8\hsize]{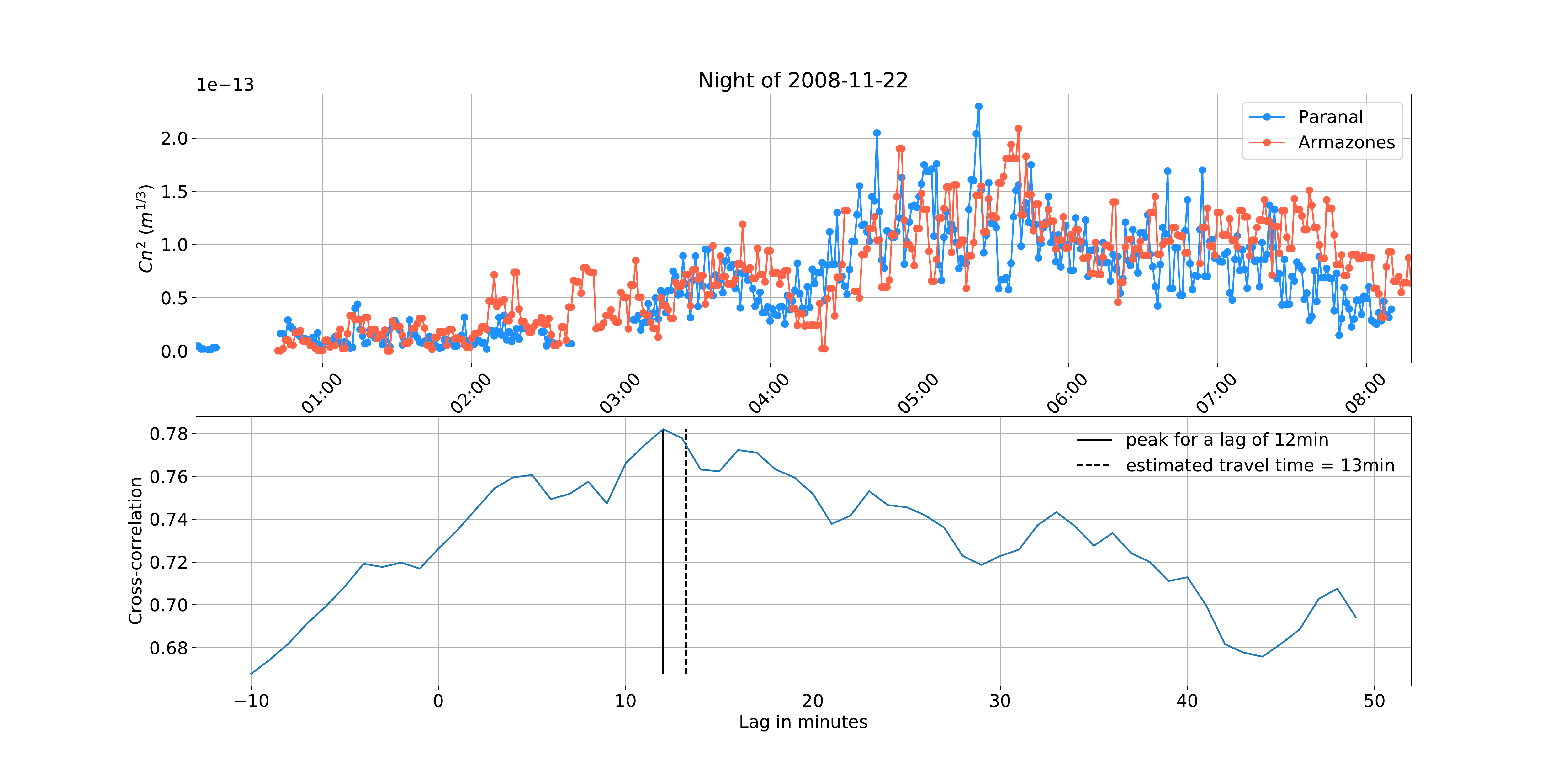}
   \includegraphics[width=0.8\hsize]{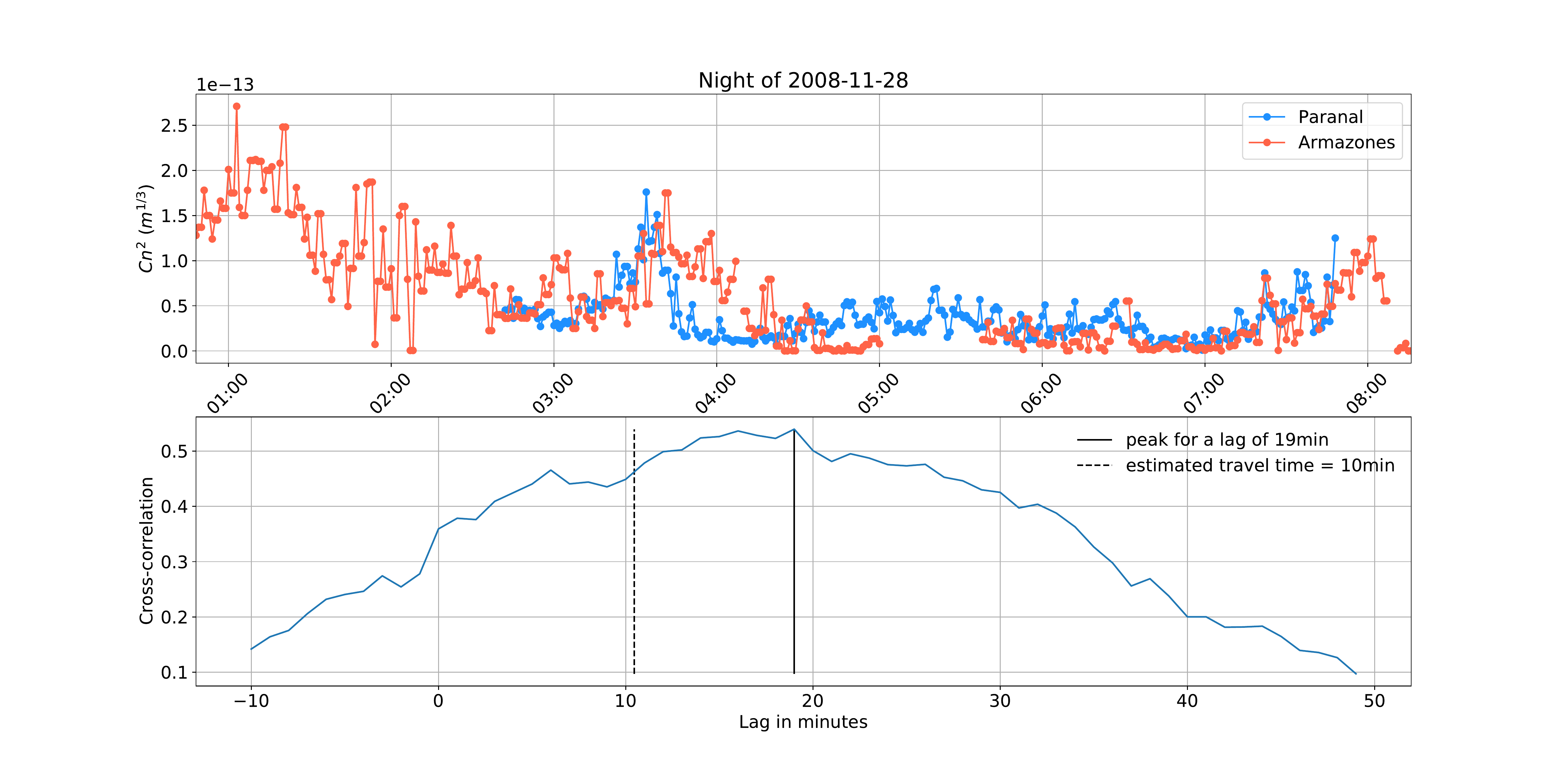}
   \includegraphics[width=0.8\hsize]{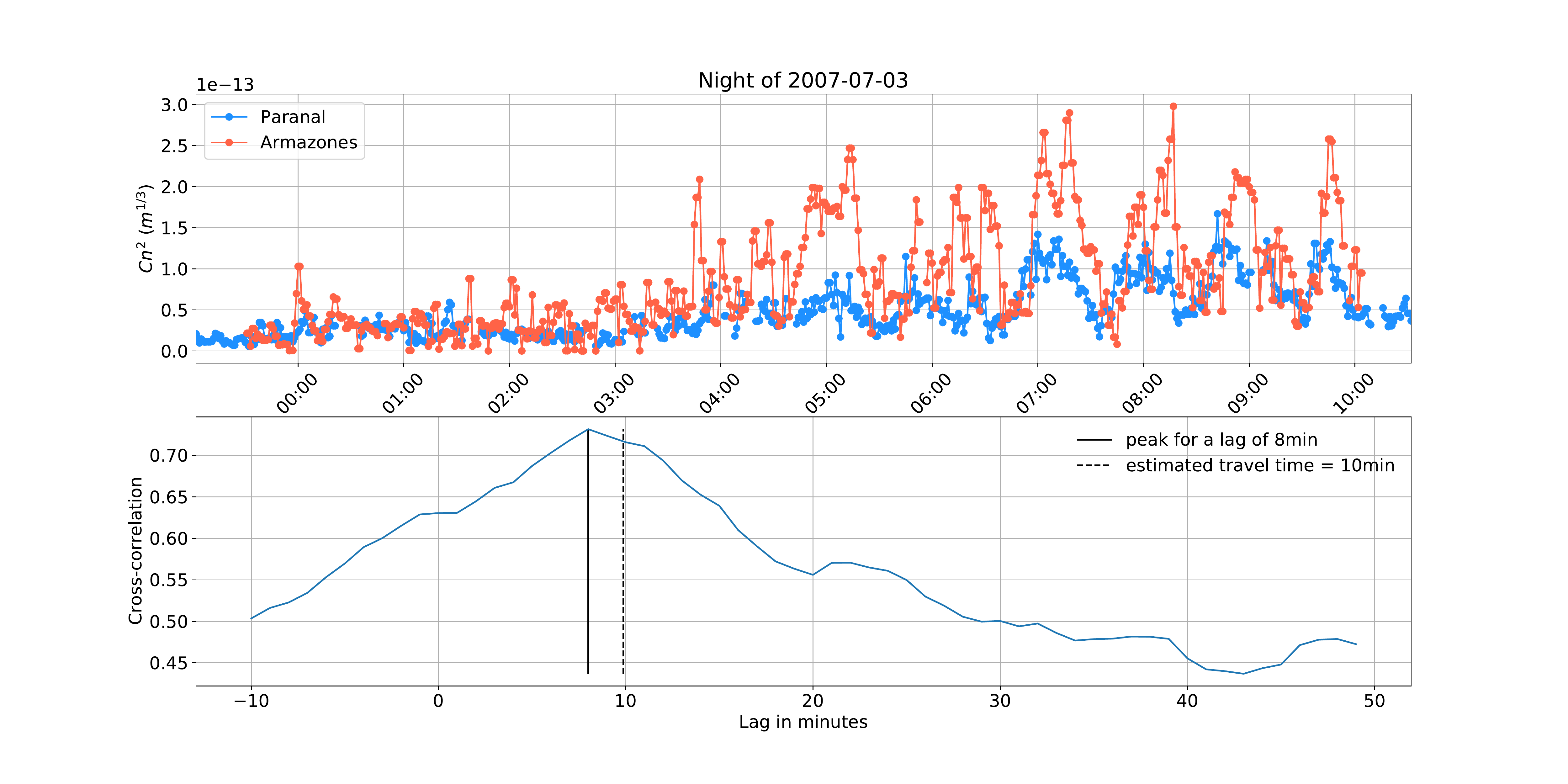}
	\end{center}
   \caption[ex] 
   { Examples of time series of the $C_n^2.dh$ measured by the MASS in the layer at 8\,km in Paranal (red) and Armazones (blue) for a few well correlated nights. The cross-correlation is shown by the blue curve as a function of the lag between both time series. The maximum cross-correlation is shown as a vertical black line, while the estimated travel time from the wind speed and direction (reanalysis ECMWF data) is shown as a black vertical dashed line.
   \label{fig_examples_CC} 
   }
   \end{figure}

We illustrate in Fig. \ref{fig_examples_CC} some examples of the $C_n^2$ times series and their cross-correlation for nights with a clear peak in the cross-correlation. In those nights, visual inspection is enough to detect the similarity between those time series. In the first two examples of Fig. \ref{fig_examples_CC}, the amplitude of the fluctuations is similar for both sites, while the third example illustrates a case with similar trends but fluctuations with a larger amplitude in Armazones compared to Paranal. 

\begin{figure} [h]
   \begin{center}
   \includegraphics[width=1.0\hsize]{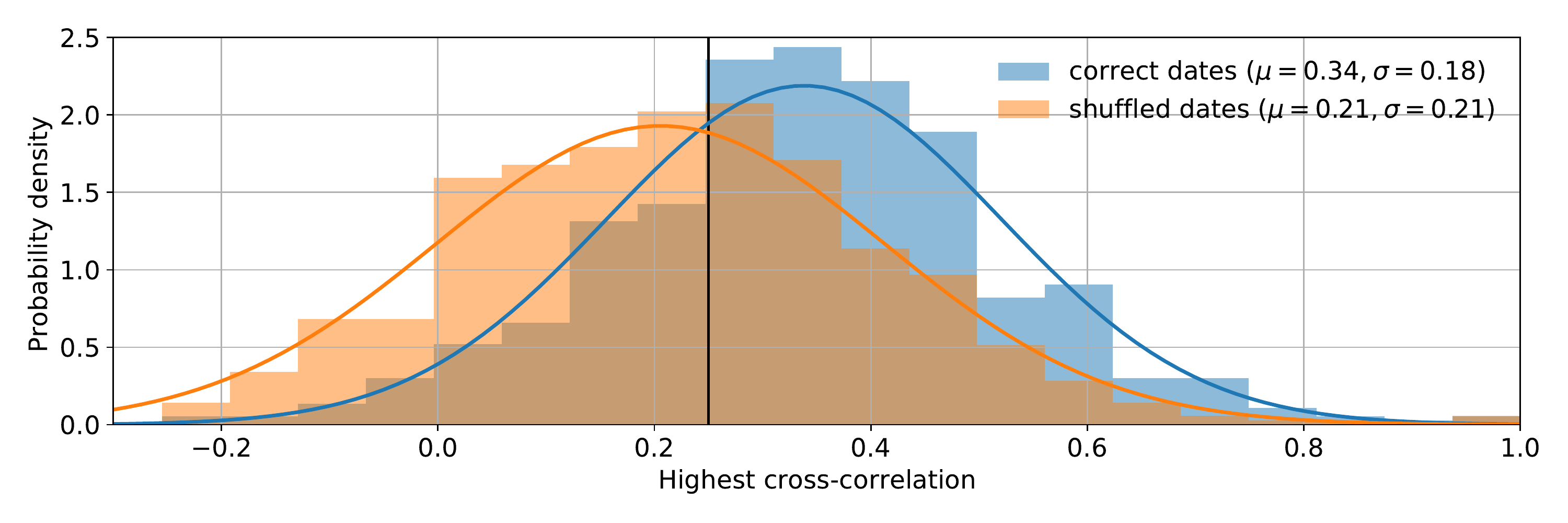}
	\end{center}
   \caption[ex] 
   {  
   Histogram of the peak cross-correlation between the MASS 8\,km layer of Armazones and Paranal. The blue histogram corresponds to the cross-correlation of time series obtained during the same night, while the orange histogram provides a reference and was obtained with Paranal and Armazones times series obtained two nights apart, where we therefore do not expect significant cross-correlation. The value of $\mu$ and $\sigma$ indicated in the label is the mean and standard deviation of the histogram, and the plain lines shows the corresponding Gaussian fit.
   \label{fig_histo_peak_cross_correlation} 
   }
   \end{figure}

To quantify how often a significant correlation in the high-altitude turbulence can be detected between Paranal and Armazones, we show in Fig. \ref{fig_histo_peak_cross_correlation} (in blue) the histogram of the maximum cross-correlation from the 571 nights. The average value is 0.34 with a standard deviation of 0.18. We also built a reference scenario where we shifted the dates of the Armazones MASS data by two days and repeated the analysis. In this scenario, we do not expect any significant correlation between the two times series, or at least not from the transport of stationary turbulence by the wind. This scenario provides therefore a reference indicating the expected distribution of the maximum cross-correlation in the absence of correlation. It is overplotted in Fig. \ref{fig_histo_peak_cross_correlation} (orange histogram): the average is 0.21 with a standard deviation of 0.21. The two distributions are statistically different: a Kolmogorov-Smirnov two-sample test gives a p-value of $2\times10^{-24}$, a value small enough to reject the hypothesis that both samples originate from the same parent population. From these two histograms, we can infer that the number of nights with a maximum cross-correlation higher than the random scenario is 28\%. We note that all these cases occur when the cross-correlation is higher than 0.25 (the vertical black line in Fig. \ref{fig_histo_peak_cross_correlation}), which happened in 71\% of the nights. 

\begin{figure} [h]
   \begin{center}
   \includegraphics[width=0.7\hsize]{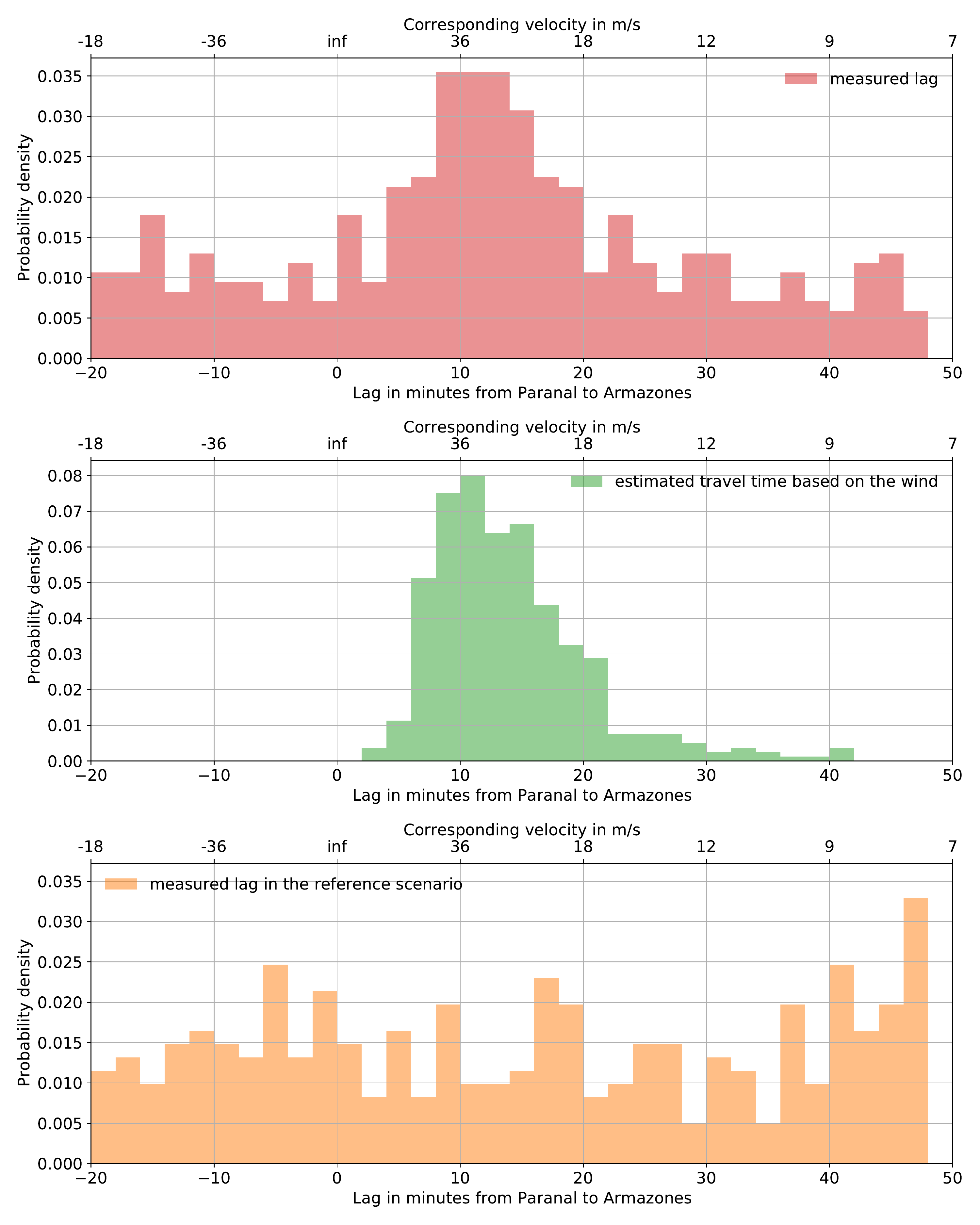}
	\end{center}
   \caption[ex] 
   {  
   Histogram of the lag  between Paranal and Armazones at the maximum cross-correlation, as measured for the 571 nights (top), as measured in the reference scenario indicating the noise level, and as estimated from the distribution of the wind speed and direction in the MASS 8\,km layer (coming from global circulation models). A positive value means the Armazones times series is delayed with respect to the Paranal times series. 
   \label{fig_histo_lag} 
   }
   \end{figure}

The histogram of the lag between Paranal and Armazones is plotted in Fig. \ref{fig_histo_lag} (top). It shows an excess of nights with a lag between 5 and 20min. This excess matches very well what is expected from the wind speed and direction (Fig. \ref{fig_histo_lag} middle), as computed based on the global circulation model reanalyses, with the bulk of the lag at about 10\,min, corresponding to a projected travel speed of 36\,m/s. In the reference scenario supposed to produce a random cross-correlation (Fig. \ref{fig_histo_lag} bottom), we indeed see a flat histogram with random fluctuations. The level of those fluctuations is similar to that seen in the real measurements for either negative lags or lags greater than 25\,min, for which we expect very few cases based on the wind distribution.
We therefore conclude that in the MASS 8\,km layer, we can detect above the noise in about 28\% of the cases, a correlation between Armazones and Paranal compatible with the transport of the turbulence by the wind. 

\section{Limitations and future prospects}

We restricted the scope of this study to the detectability of a correlation between Paranal and Armazones. It now remains to be investigated how to best exploit this information in order to predict the turbulence. Given the overall wind directions in high altitude, we can hope to anticipate the turbulence in Armazones based on the measurements done at Paranal. As shown in Fig. \ref{fig_examples_CC} bottom, the turbulence might have developed during the travel time between the two sites. Therefore the knowledge of the maximum cross-correlation and the time lag alone is not enough to be able to predict the future, but it is a significant step forward to anticipate sudden changes in the turbulence, which was identified as the weakness of previous nowcast systems. 

We considered in this study only the high-altitude turbulence located between 4 and 12\,km, and centered at 8\,km, because this was the most promising layer to look for correlations between Paranal and Armazones. A similar analysis can be done for the other high-altitude turbulent layers. As the wind is slower in those layers (Fig. \ref{fig_wind_rose_MASS_layers}), we expect to find longer time lags, and maybe an overall decrease in the maximum cross-correlation because the turbulence had more time to evolve. 

The Armazones and Paranal data are likely not relevant to investigate the transport of the turbulence in the ground layer and in lower altitude layers. First, there is a significant difference in altitude between these two summits, which means that the two MASS response functions to the lowest layer (500m) do not overlap. Second, the ground layer and 500\,m layer are dominated by northern winds, meaning it is less likely to detect a correlation unless the fronts of turbulent air are coherent over a width larger than the Paranal-Armazones distance (22km). To predict ground layer turbulence, monitors located upfront, in the North of Cerro Paranal or Cerro Armazones would be better suited. This is the case of Cerro Ventarrones located about 20km North of Cerro Armazones, for which archival turbulence data already exists. 

The MASS instruments themselves are also a potential source of noise. First, the Paranal and Armazones MASS instruments are not identical. Their response function is likely slightly different, which can dilute the cross-correlation. Second, they are operated independently. They  were likely pointing at different stars at any given time. For instance, a sensor looking at a star with a zenith angle of $20^\circ$ probing a layer at 8\,km altitude is measuring the turbulence 2.9\,km away from the sensor location. This means that the direction of the turbulence transport probed by the cross-correlation technique is not exactly the Paranal Armazones direction but can be offset by up to 10 degrees. In addition the brightness and the properties of the star used to estimate the turbulence likely affects the signal-to-noise of the $C_n^2$ estimation.  It would therefore be interesting to repeat those measurements with identical MASS sensors synchronised between the two sites and looking at identical stars.

The turbulence in Paranal and Armazones is dominated by the ground layer. This means that nowcast systems must include an accurate treatment of the ground layer. The wind speeds are typically smaller than 10m/s in this layer and the topography of the Andes disturbs the flow much more than in higher altitudes. For this reason, a suite of turbulence monitors located within a few kilometres of Paranal or Armazones are probably best suited to nowcast the boundary layer conditions. The use of mesoscale models to optimise the location of those sensors is an interesting prospect in this respect. A similar cross-correlation analysis as presented in this study could be done on synthetic data to validate the best sensor locations.

\section{Conclusions} 

After reminding the motivations, accuracy and limitations of current techniques to nowcast the seeing, we presented an approach using sensors located at different sites to anticipate the optical turbulence. We validated this approach by studying the cross-correlation between the turbulence in a high-altitude layer above Paranal and Armazones, as described by the MASS 8\,km $C_n^2$ values. We showed that we can detect in about 28\% of the cases a correlation between Armazones and Paranal, with a lag compatible with the transport of the turbulence by the wind. In the other cases, there is either no significant correlation, or a correlation to a level which is not detectable in the dataset used in this work, which was collected to provide statistical parameters for each site and not tailored for cross-correlation measurements. Identical turbulence profilers looking at the same stars and synchronised could validate such results and decrease the noise in the cross-correlation estimation. It could be implemented in the context of the Astronomical Site Monitoring system that is planned for the ELT at Cerro Armazones. 
These results are encouraging, because they suggest that we can anticipate the turbulence, if the appropriate turbulence sensor is located upstream of the site.

\acknowledgments 
We thank R. Querel for valuable discussion on turbulence nowcasting and on the analysis of seeing fluctuations. We thank Aleksandra Solarz for sharing her expertise in Detrended Fluctuation Analysis (DFA) applied to the Paranal seeing measurements. We acknowledge the Copernicus Climate Change Service (C3S) which was used to access the fifth generation ECMWF reanalysis for the global climate and weather for the past 4 to 7 decades\cite{ERA5} .

\bibliography{biblio} 
\bibliographystyle{spiebib} 

\end{document}